%% file: main_arXiv.tex
\documentclass[12pt]{article}

\usepackage[utf8]{inputenc}
\usepackage{amsmath}
\usepackage{amsthm}
\usepackage{amsfonts}
\usepackage{mathtools}
\usepackage{amssymb}
\usepackage{comment}
\usepackage{algorithm}
\usepackage{algpseudocode}

\usepackage{etoolbox}

\usepackage{dsfont}
\usepackage{stackengine}
 \usepackage{booktabs}
 \usepackage{enumitem}
  \usepackage{pdfpages}

\usepackage[hidelinks]{hyperref}

\usepackage{tikz}
\usepackage{dblfloatfix} % fix for bottom-placement of figure
\usepackage[labelfont=bf]{caption}
\usepackage{subcaption}
\usepackage{tkz-tab}
\usepackage{dsfont}
\usepackage{wrapfig}
\usepackage{wasysym}
\usepackage{enumitem}
\usepackage{adjustbox}
\usepackage{ragged2e}
\usepackage{longtable}
\usepackage{changepage}
\usepackage{setspace}
\usepackage{hhline}
\usepackage{multicol}
\usepackage{tabto}
\usepackage{float}
\usepackage{multirow}
\usepackage{makecell}
\usepackage{fancyhdr}
\usepackage{mathdots}
\usepackage{graphicx}
\usepackage{verbatim}
\usepackage{times}

\usepackage{algorithm}%http://ctan.org/pkg/algorithms
\usepackage{algpseudocode}%

\usepackage{lettrine}

\newenvironment{sciabstract}{%
\begin{quote} \bf}
{\end{quote}}

\usepackage{dsfont}
\usepackage{amsfonts}
\usepackage{amsthm}
\usepackage{amsmath}
\usepackage{amscd}
\usepackage{amssymb}
\usepackage{enumitem}

\usepackage{float}

\usepackage{caption}
\usepackage{subcaption}

\usepackage{comment}

\theoremstyle{definition}

\newtheorem{?}[Th]{Problem}

\newcommand{\ZLnote}[1]{\textcolor{red}{\textbf{#1}}}

\usepackage{mathtools}
\mathtoolsset{showonlyrefs}
\usepackage[normalem]{ulem}

\usepackage{wasysym}
\usepackage{yhmath}

\title{\LARGE{Social learning moderates the tradeoffs between efficiency, stability, and equity in group foraging}}

\author{Zexu Li$^{a}$, M. Amin Rahimian$^{b,\ast}$ and Lei Fang$^{a,c,d\ast}$ \\
\normalsize{$^a$~Civil and Environmental Engineering, University of Pittsburgh \vspace{-1mm}}\\
\normalsize{$^b$~Department of Industrial Engineering, University of Pittsburgh \vspace{-1mm}}\\
\normalsize{$^c$~Department of Mechanical Engineering and Materials Science, University of Pittsburgh \vspace{-1mm}}\\
\normalsize{$^d$~Department of Bioengineering, University of Pittsburgh \vspace{-1mm}}\\
\normalsize{$^\ast$~To whom correspondence should be addressed; email: \{rahimian,lei.fang\}@pitt.edu.}
}

\date{}
\begin{document}

\maketitle

\vspace{-10mm}

\begin{sciabstract}

\input{sections/abstract_new}

\end{sciabstract}

\input{sections/body_new}

\section*{Methods} 

\input{sections/matmethods}

% \section*{Data availability}

% The email dataset is publicly available at \cite{yin2017local}. The workplace, conference, and hospital datasets are publicly available at \cite{genois2018can} and \cite{vanhems2013estimating}, and the Utah school dataset is available at \cite{toth2015role}.

% \section*{Code availability}
% Code for reported simulations can be accessed from \href{https://github.com/shamsvahid2/topology-contagion}{https://github.com/shamsvahid2/topology-contagion}.

\section*{Acknowledgements}{}
\input{sections/acknowledgement}

\bibliographystyle{Science}

\bibliography{levy-refs}

\input{sections/si}

\end{document}

%% file: sections/abstract_new.tex
Collective foragers—from animals to robotic swarms—must balance exploration and exploitation to locate sparse resources efficiently. While social learning is known to facilitate this balance, how the range of information sharing shapes group-level outcomes remains unclear. Here, we develop a minimal collective foraging model in which individuals combine independent exploration, local exploitation, and socially guided movement. We show that foraging efficiency is maximized at an intermediate social learning range, where groups exploit discovered resources without suppressing independent discovery. This optimal regime also minimizes temporal burstiness in resource intake, reducing starvation risk. Increasing social learning range further improves equity among individuals but degrades efficiency through redundant exploitation. Introducing risky (negative) targets shifts the optimal range upward; in contrast, when penalties are ignored, randomly distributed negative cues can further enhance efficiency by constraining unproductive exploration. Together, these results reveal how local information rules regulate a fundamental trade-off between efficiency, stability, and equity, providing design principles for biological foraging systems and engineered collectives.

%% file: sections/body_new.tex
\noindent Efficient resource acquisition is a critical challenge for mobile foragers—including albatrosses \cite{viswanathan1996levy}, marine predators \cite{sims2008scaling,sims2012levy}, and human hunter-gatherers \cite{raichlen2014evidence,venkataraman2017hunter,pacheco2019nahua}—who must locate spatially sparse and unpredictable food while expending limited energy and avoiding hazards. These constraints motivate strategies that balance wide-ranging exploration with localized exploitation.

A foundational framework for describing animal movement is the L\`evy walk \cite{viswanathan1996levy,sims2008scaling,sims2012levy}, in which step lengths follow a power-law distribution $P(l)\sim l^{-\mu}$ ($1< \mu\leq 3$), generating clusters of short displacements punctuated by rare long relocations. Although sometimes conflated with L\`evy flights—whose instantaneous steps lack physical realism—L\`evy walks capture finite-speed motion and have been supported empirically across diverse systems, from wandering albatrosses \cite{viswanathan1996levy} to T cells \cite{harris2012generalized}, swarming bacteria \cite{ariel2015swarming,ariel2017chaotic}, honeybees \cite{reynolds2007displaced}, marine predators \cite{sims2008scaling,sims2012levy}, and human foragers \cite{raichlen2014evidence}. Early claims of L\`evy scaling were tempered by statistical reanalyses \cite{edwards2007revisiting}, but subsequent improved methodology reaffirmed its relevance. While L\`evy strategies enhance encounters with sparse targets \cite{viswanathan1999optimizing,viswanathan2008levy,humphries2014optimal}, real foragers rarely follow memoryless step-length statistics. Upon finding prey, individuals typically switch to area-restricted search (ARS) characterized by slow, tortuous movements near resource patches \cite{ross2018general,benichou2006two,benhamou2015ultimate,pyke2015understanding}. ARS reduces search time \cite{benichou2006two}, increases encounter rates, and lowers starvation risk \cite{ross2018general}, especially when resources are patchy.

However, many organisms do not forage alone. When unsuccessful individuals encounter other individuals exploiting rich patches, they can abandon independent search and join them. Such collective search is widespread across taxa \cite{barnard1981producers,detrain2008collective,ginelli2015intermittent,gomez2022intermittent} and is especially advantageous when resources are scarce, clustered, or ephemeral \cite{giraldeau2018social}. The efficiency of collective problem solving depends critically on how information flows through social networks. Network-theoretic analyses show that topology regulates the exploration–exploitation balance \cite{lazer2007network}; experiments demonstrate that communication networks can enhance group learning \cite{mason2012collaborative}; and social-influence structure shapes collective accuracy \cite{becker2017network}. In this context, the communication radius 
$\rho$ can be viewed as a spatial analogue of network connectivity: small 
$\rho$ encourages independent search, while large $\rho$ increases conformity to shared discoveries. By linking ecological foraging models to social-network science, we provide a mechanistic perspective on how connectivity regulates group search.

At the core of these processes is social learning—the acquisition and use of information from others \cite{bhattacharya2014collective,heins2024collective}. Social learning influences resource acquisition by tuning how individuals respond to peer discoveries. Empirical work on Mongolian gazelles shows that intermediate-distance vocal communication maximizes patch detection while avoiding over-aggregation \cite{martinez2013optimizing}. Computational studies demonstrate that probabilistic copying can generate L\`evy-like statistics \cite{bhattacharya2014collective}, while theory reveals two fundamental trade-offs—exploration vs.\ exploitation and individual vs.\ social search—that govern efficiency \cite{garg2022individual}. Evolutionary models further show that social learning promotes coexistence of diverse strategies, although excessive copying can harm group performance unless moderated by heuristics such as ARS \cite{garg2024evolution}. {Understanding these mechanisms is relevant not only for animal foraging but also for engineered multi-agent systems—such as swarm robotics and distributed sensing—where communication range is a controllable design parameter.}

Despite these advances, a mechanistic understanding of how information-sharing range determines group-level outcomes remains lacking. {The central goal of this study is to understand how social learning range shapes collective foraging. Specifically}, we address the following research questions (RQs):
\begin{quote}
RQ1. How does the social learning parameter, $\rho$, mediate the optimal resolution of the tradeoff between exploration and exploitation in collective search, and how does the optimum $\rho$ vary with factors such as target distribution and group size? \\ 
% How do environmental factors and group search characteristics such as distribution of targets and number of agents affect the choice of optimum $\rho$?\\
RQ2. How does $\rho$ influence other collective characteristics such as temporal stability and equity of resources found among individuals, as well as their network connectivity?
\end{quote}

% \LFnote{What about negative target?}

% \sout{Here, we introduce a deterministic, range-based social-learning model integrated with ARS to investigate how communication distance $\rho$ regulates collective foraging. Agents alternate among three behaviors—exploration, exploitation, and targeted walk—allowing us to quantify how $\rho$ shapes mean efficiency, temporal variability, and equity in resource distribution. We further incorporate negative targets to evaluate how social learning enables risk mitigation.} 

{Here, we develop a collective foraging model that integrates ARS with range-limited social learning. Individuals switch among three behavioral modes—exploration, exploitation, and targeted movement—regulated by a single communication-distance parameter $\rho$, which sets how far social signals propagate. This formulation allows us to quantify how $\rho$ jointly shapes (i) mean search efficiency, (ii) the temporal stability and equity of resource intake, (iii) network connectivity among agents, and the emergent interaction structure underlying these outcomes. We then extend the same framework to risky environments by introducing negative targets that generate avoidance cues shared through the same social-learning rule.}

% \sout{Our results reveal that $\rho$ governs a fundamental three-way trade-off: as 
% $\rho\rightarrow 0$, agents act independently, maximizing exploration; intermediate $\rho$ balances exploration and exploitation to maximize efficiency; and large $\rho$ maximizes equity but reduces efficiency and induces bursty temporal dynamics. Strikingly, we find that introducing “fake’’ negative cues—signals that mark resource-poor regions—can enhance collective efficiency by steering exploration away from unproductive areas. This counterintuitive result suggests that imperfect or even misleading information can improve group performance, paralleling effects seen in other collective systems \cite{koster2022spurious}. Together, these results provide a minimal mechanistic framework explaining how social-communication range shapes collective foraging performance, offering insight into natural systems and providing design principles for artificial swarms.}

{Our results show that $\rho$ induces a non-monotonic performance landscape: small $\rho$ preserves independent exploration but limits coordination, intermediate $\rho$ balances exploration and exploitation to maximize efficiency, and large $\rho$ promotes homogeneous behavior that increases equity but reduces efficiency and produces bursty intake dynamics. Mechanistically, this optimum coincides with an intermediate level of emergent connectivity in the proximity network induced by social learning. Additionally, when negative targets are introduced, the optimal $\rho$ shifts toward larger values as avoidance becomes more valuable; moreover, when penalties are ignored, randomly distributed negative cues can increase efficiency by pruning unproductive search regions. Together, these findings provide a minimal mechanistic framework for how information-sharing range shapes collective foraging, with implications for biological groups and engineered swarms.}

\begin{figure}[h!]
\centering
\includegraphics[width=.7\linewidth]{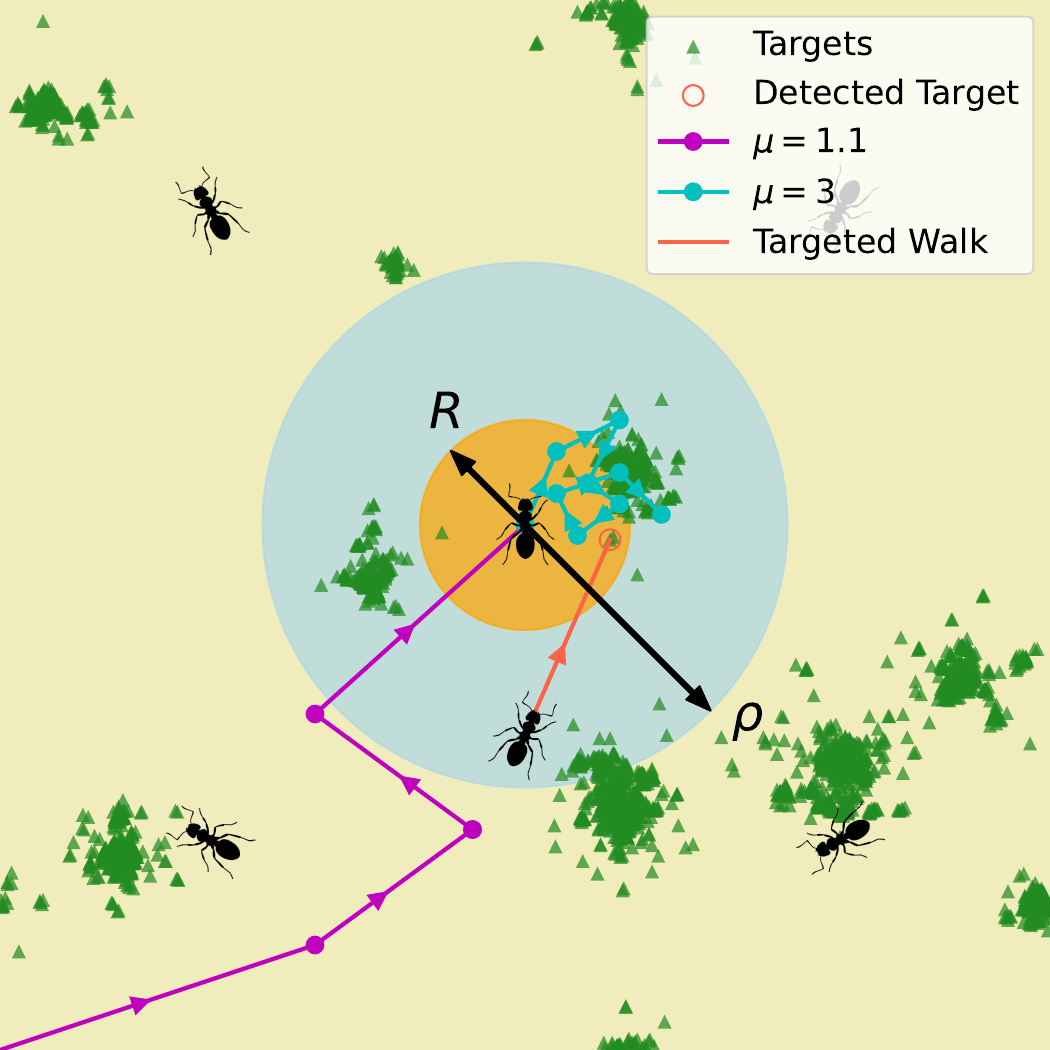}
\caption{\textbf{A schematic of the collective search model.} Agents begin by performing independent L\`evy walks with $\mu = 1.1$ (exploration, shown in purple) until they detect a target. Upon detection, the agent ceases social learning and switches to a L\`evy walk with $\mu = 3$ (exploitation, shown in blue) within a circular region of radius $R$ centered at the detected target, which is updated over time. Meanwhile, a social signal emitted by this agent attracts other agents within a region of radius $\rho$ to perform targeted walk towards the detected target (shown in orange). Agents outside this region do not receive the social signal and maintain their search mode.}
\label{fig:1}
\end{figure}

\section*{Results}
% \sout{\ZLnote{We first examine how social learning range affects collective efficiency, then analyze its impact on temporal stability, equity, network structure, and risk-sensitive foraging. To this end,} we simulate our group foraging model by varying the social learning range from $\rho= 0$ to $\rho = 0.7$ \ZLnote{--the largest value permitted by the domain--under} \sout{, which is the largest social learning range in our domain, using a }periodic boundary condition. \sout{Considering that the targets are limited and nonregenerative, search efficiency will definitely drop as more targets are depleted. Each simulation will end when $30\%$ of the targets have been collected}. We also vary the number of patches, keeping the total number of targets across all patches fixed at $6000$. Furthermore, we introduce negative targets, conceptualized as risks encountered by agents during the search, that negatively impact search efficiency. When an agent encounters a negative target, it can also send a social signal to inform agents within a radius $\rho$ of the presence of the negative target. Upon approaching known negative targets, agents exhibit active avoidance behaviors. } 

{We organize the Results around the two research questions posed in the Introduction.  We first address \textbf{RQ1} by quantifying how the social-learning range $\rho$ mediates the exploration--exploitation trade-off in collective search and produces an intermediate $\rho$ that maximizes collective efficiency. We then address \textbf{RQ2} by characterizing how $\rho$ reshapes collective properties beyond mean efficiency---including temporal stability of target intake, equity of individual rewards, and the emergent proximity-network connectivity that underlies information flow and coordination. Finally, building on the mechanisms identified in \textbf{RQ1} and {RQ2}, we present a post-hoc application in which the same framework is extended to a risky environment (containing negative targets), illustrating how the role of $\rho$ generalizes under additional task constraints. To this end, we simulate the group foraging model on a periodic domain while sweeping $\rho$ from $0$ to $0.7$ (the maximum feasible range in our geometry). We vary the number of patches while holding the total number of positive targets fixed at $6000$. To model risky environments we introduce negative targets that reduce efficiency; upon encountering a negative target, an agent broadcasts its location within radius $\rho$, and informed agents exhibit active avoidance behavior.} Further details of the simulation setup are provided in Methods.

\subsection*{Efficiency} 
\begin{figure*}[h!]%[tbhp]
\centering
\includegraphics[width=0.61\textheight]{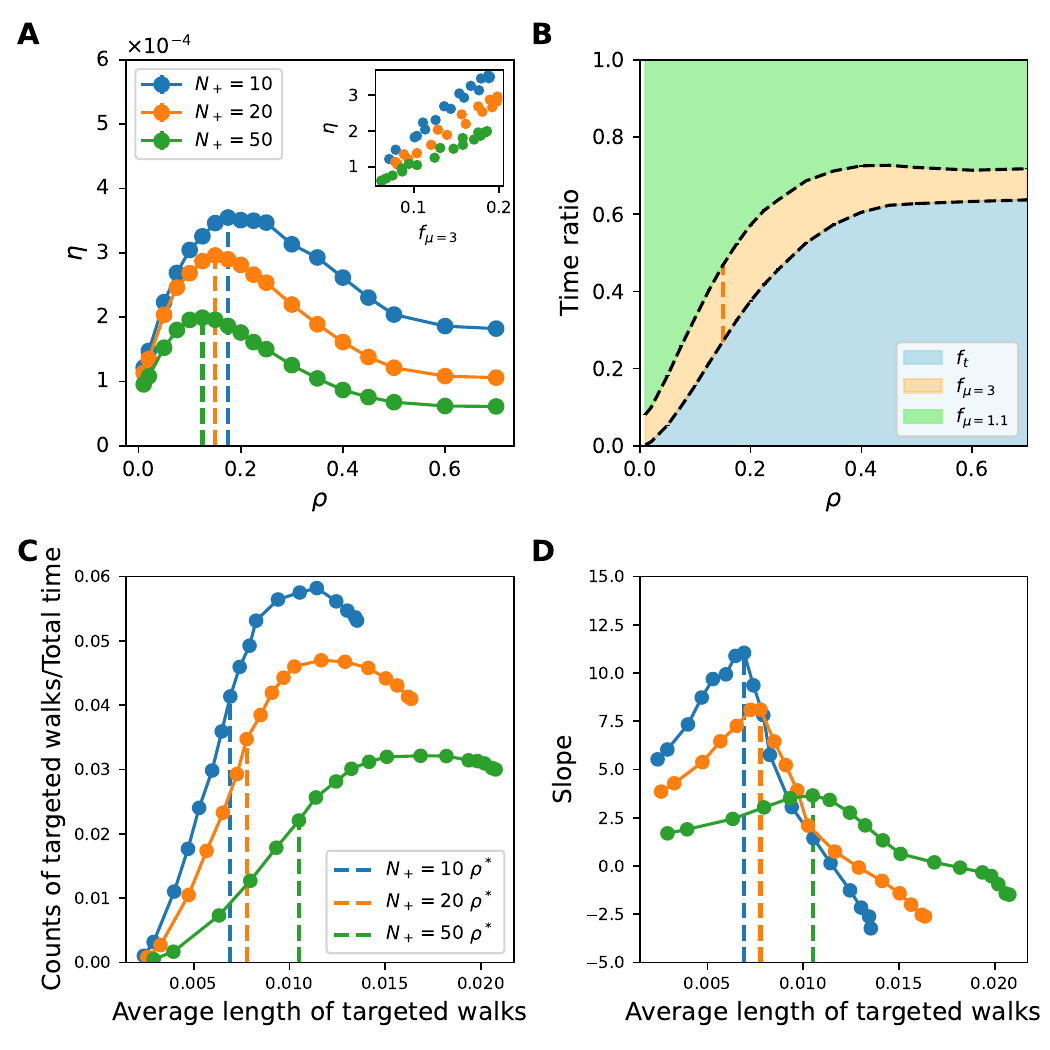}
\caption{\textbf{Optimal $\rho$ mediates the tradeoff between exploration and exploitation.} (\textit{A}) Efficiency versus $\rho$ for a fixed number of agents ($N_A=50$) and three numbers of positive patches ($N_+=10;20;50$). The total number of targets is fixed at 6000 with $\beta=2$. Error bars denote 95\% confidence intervals but are too small to be visible. The inset shows that efficiency increases monotonically with the $f_{\mu=3}$. (\textit{B}) Average time fractions spent in three behaviors for the $N_+=10$ case in (\textit{A}); the optimal $\rho$ aligns with the social-learning regime where agents spend the largest fraction of time exploiting targets ($\mu=3$). (\textit{C}) Counts of targeted walks, normalized by total time (benefit), plotted against their average length (cost). (\textit{D}) Slope of the curve in (\textit{C}) as a function of targeted-walk length, showing that the highest benefit-to-cost ratio coincides with the optimal $\rho$. \label{fig:2}} %
\end{figure*}

The group efficiency, defined as $\eta=N_{collected}/N_{positive}T$, quantifies the rate at which agents collect positive targets. Here, $N_{collected}$ and $N_{positive}$ are the number of targets collected by agents and the total number of targets within the domain, respectively. $T$ is the total search time. As the social learning range ($\rho$) increases, it first increases and then decreases, with an optimal $\rho$ that yields the highest efficiency. (Fig. \ref{fig:2}\textit{A}). This phenomenon is robust across different numbers of patches, whether we fix the total number of targets or the number of targets per patch (see Supplementary Fig. \ref{Supplfig:4}). It can also be observed for target distributions with $\beta = 1.1$ (see Supplementary Fig. \ref{Supplfig:2}) and $\beta = 3$ (see Supplementary Fig. \ref{Supplfig:3}), as well as different group sizes when $N = 10$ (see Supplementary Fig. \ref{Supplfig:5}). {This non-monotonic dependence is consistent with previously identified trade-offs between exploration and exploitation and between individual search and social learning \cite{garg2022individual}.} {Here, our explicit definition of exploration and exploitation provides a clearer mechanistic account of the emergence of an optimal $\rho$.} 

%Variation of efficiency ($\eta$) and time ratios of agents' behaviors, exploration ($f_{\mu=1.1}$), exploitation ($f_{\mu=3}$), and targeted walk ($f_t$), with social learning range ($\rho$) for a hierarchical target distribution with $\beta=2$. 

Agents in our model exhibit three modes: L\`evy walk with $\mu=1.1$, L\`evy walk with $\mu=3$, and targeted walks triggered by social learning. We interpret $\mu = 1.1$ mode as global exploration, $\mu = 3$ mode as area restricted exploitation, and the targeted walk as the operational cost associated with exploitation. Agents can adaptively switch between the three modes as needed.  We measure the time spent on each mode, as shown in Fig. \ref{fig:2}\textit{B}. When $\rho=0$ (no social learning), targeted walks do not occur, and agents search independently, spending approximately 90\% of their time on exploration. After encountering a target, an agent shifts from exploration to exploitation, 
%to increase the encounter rate with targets
which contributes to the remaining 10\% of time spent on exploitation. As $\rho$ increases, targeted walks become more prevalent, triggering greater exploitation. Consequently, the time ratio of exploration decreases as $\rho$ rises. At the optimal $\rho$, the time ratio of exploitation reaches its maximum. This reflects an elegant balance between exploration and exploitation: enough exploitation sustains high current efficiency, while just enough exploration continually discovers new patches, preserving future opportunities and, in turn, overall efficiency.
As $\rho$ increases further, targeted walks become more prevalent. However, the time ratio of exploitation decreases, indicating that an excessively high proportion of targeted walks leads to redundant search effort and reduces the exploitation ratio. This occurs because the number of targets is limited in each patch. When social signals attract most of the agents to a detected target, early-arriving agents can still collect targets, whereas late-arriving agents who travel for extended periods can only collect a few targets. Additionally, the remaining targets may be insufficient to support sustainable exploitation, causing late-arriving agents to revert to exploration soon. In other words, the social signal breaks the exploration of those late-arriving agents (far from the social signal) and lets them restart exploration from a region where most of the targets are consumed. Thus, an excessively large $\rho$, causes a high proportion of targeted walks, raises the cost and reduces the benefits of exploitation, disrupts the exploration process, and forces agents to explore regions where most targets are already consumed, thereby diminishing efficiency.   %\ZLnote{Lei: specifically mention the cost of exploit}

From Fig. \ref{fig:2}\textit{A}, we observe that the optimal $\rho$ increases as the number of patches, $N_+$, decreases, as exploiting a single patch becomes more critical than exploring new ones in this scenario. Conversely, when targets are distributed across a larger number of patches, agents must identify more patches to collect {more} targets. Additionally, efficiency is higher in scenarios with fewer patches, as agents spend less time searching for new patches. Furthermore, we plot efficiency, $\eta$, as a function of $f_{\mu=3}$ in the inset of Fig. \ref{fig:2}\textit{A}, demonstrating that $\eta$ is positively correlated with $f_{\mu=3}$. The slopes of the $\eta$--$f_{\mu=3}$ curves vary due to differences in the number of targets per patch. A larger number of targets per patch (the case with $N_+=10$) leads to a larger slope of $\eta-f_{\mu=3}$, meaning that increasing the time ratio of exploitation becomes more beneficial.

To investigate the role of social learning and the resulting targeted walk, we construct a phase diagram in \ref{fig:2}\textit{C}, plotting the counts of targeted walks normalized to the total time in each case, against the averaged length of targeted walks. In this diagram, a higher count of targeted walks signifies increased interaction with social signals, which promotes exploitation and enhances efficiency. Conversely, a larger average length of targeted walks represents high cost, as it is the distance agents must travel before they can begin detecting a resource, which reduces overall efficiency. The slope of the curves in this phase diagram, therefore, represents the ratio of benefits to costs, specifically the gain from increased exploitation per unit of cost. Based on this, we hypothesize that the point where the slope of each curve reaches its maximum corresponds to the most efficient scenario. At this point, the benefit gained from targeted walks increases most rapidly relative to their cost. To verify this hypothesis, we compute the derivative of the curves from the phase diagram as shown in Fig. \ref{fig:2}\textit{D}, which reveals the slope. From the figure, a maximum value of the derivative represents the most efficient scenario. This derivative is smoothed using a Gaussian filter with $\sigma = 1$ to reduce the noise. {Together, these results answer RQ1 by demonstrating that performance is non-monotonic in $\rho$: intermediate-range social learning maximizes efficiency by improving exploitation without fully suppressing independent discovery.}

\begin{figure*}[h!]%[tbhp]
\centering
\includegraphics[width=0.65\linewidth]{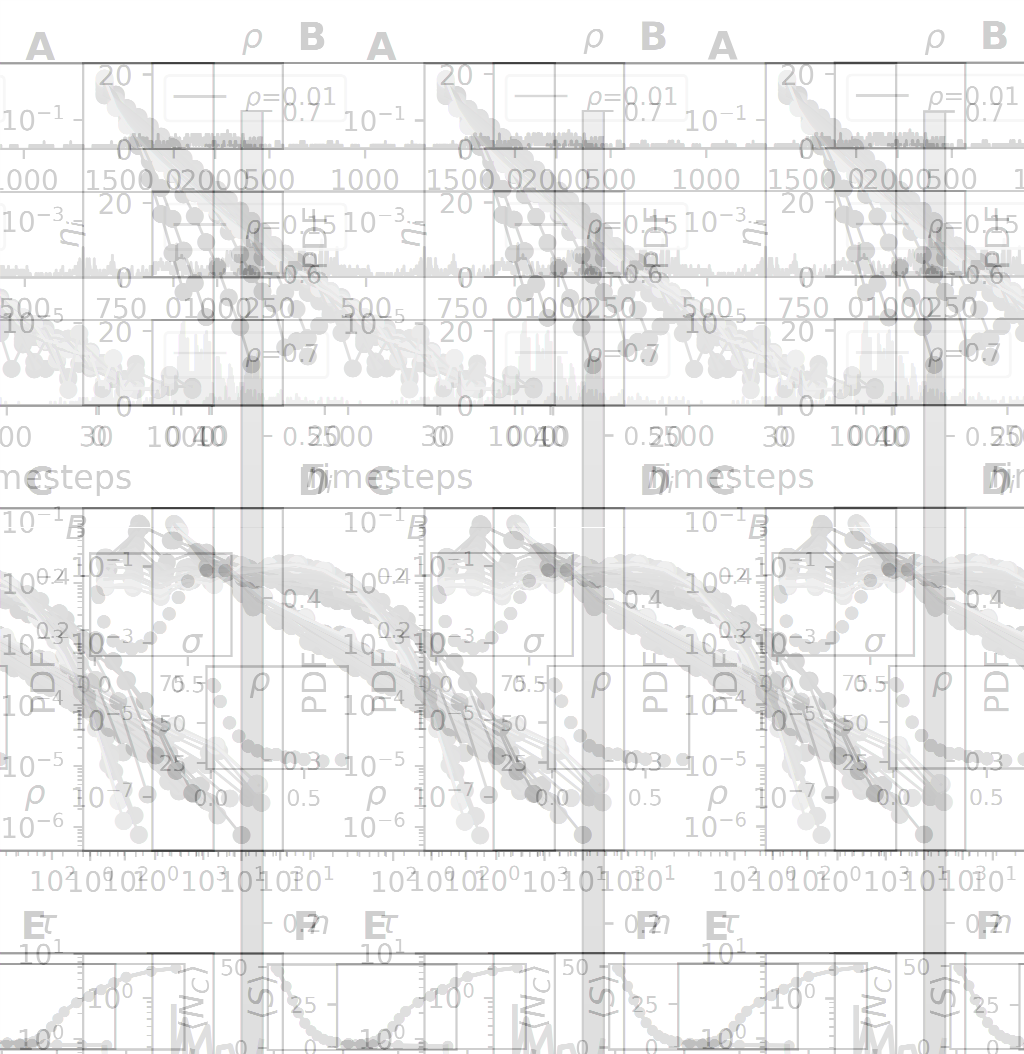}
\caption{\textbf{{Collective search characteristics induced by $\rho$: burstiness, equity, and network connectivity.}} (\textit{A}) Time variation of instantaneous efficiency $\eta_i$ with $\rho=0.01$, $\rho = 0.15$, and $\rho = 0.7$ (showing one case). $\eta_i$ represents the number of collected targets per time step.  (\textit{B}) Probability density function (PDF) of $\eta_i$ indicating that low $\rho\leq 0.2$ limits $\eta_i$ due to less social learning, while higher $\rho$ ($0.2<\rho<0.5$) allows $\eta_i$ to reach greater values with an exponential distribution. For $\rho\geq 0.5$, the proportion of medium $\eta_i$ values decreases, though $\eta_i$ can achieve higher values with a heavy-tail distribution. (\textit{C}) PDF of inter-event time, defining an event as when the instantaneous efficiency $\eta_i\geq 1$. The inset figure illustrates the variation of the burstiness parameter $B$ with $\rho$. The optimal $\rho$ corresponds to the lowest B, indicating a steady intake of resources and, consequently, a minimal starvation time. (\textit{D}) PDF of number of collected targets by each agent, with an inset indicating that the standard deviation (representing equity) decreases as $\rho$ increases. (\textit{E}) PDF of the average component size in each case $S$. The PDF of $S$ shifts to the right as $\rho$ increases, indicating that the case-averaged component size $\left\langle S \right\rangle$ increases with $\rho$, as shown in the inset. (\textit{F}) PDF of average number of components in each case $N_C$. The PDF of $N_C$ shifts to the left as $\rho$ increases, indicating the case-averaged number of components $\left\langle N_C \right\rangle$ decreases with increasing $\rho$, as shown in the inset. All data is based on a scenario with 50 agents and $N_+=10$.
\label{fig:3}} 
\end{figure*}

\subsubsection*{Timeseries characteristic and equality} We select $\rho = 0.01$, $\rho = 0.2$, and $\rho = 0.7$ to illustrate the temporal variation in the number of targets collected by all agents per time step (Fig. \ref{fig:3}\textit{A}), which we define as instantaneous efficiency, $\eta_i$. Here, the number of agents is set at $N_A = 50$, and the domain contains 10 patches, each with 600 targets. When $\rho = 0.01$, agents search nearly independently and share minimal information about target locations, resulting in a limited $\eta_i$ that remains low for an extended period. This occurs because the number of agents visiting a patch limits the number of collected targets per timestep. As $\rho$ increases to $\rho = 0.15$ (the optimal $\rho$), $\eta_i$ reaches higher values, indicating agents can cooperate effectively to collect targets in one patch. 
%Moreover, the efficiency maintains a long time with $\eta_i>0$. 
As $\rho$ increases further, such as $\rho = 0.7$, targeted walks attract more agents to exploit the same patch, which increases instantaneous $\eta_i$ further. However, this high efficiency cannot be sustained, as excessive agent exploitation rapidly depletes resources, leaving little exploration. As a consequence, $\eta_i$ rapidly decays to a low level. Upon the total consumption of the existing target, agents require time to explore new target patches, during which $\eta_i$ drops to zero. Thus, a larger $\rho$ results in a bursty instantaneous efficiency. 

To highlight the burstiness of $\eta_i$, we plot its probability density functions (PDFs), as shown in Fig. \ref{fig:3}\textit{B}. The PDF reveals that a larger $\rho$ increases the proportion of higher $\eta_i$ values, resulting from social learning and agent aggregation. To further reveal the bursty pattern existing in the timeseries of instantaneous efficiency, we use the burstiness parameter introduced by Goh and Barab\'asi \cite{goh2008burstiness}. Here, we defined an event as occurring when $\eta_i$ exceeds 0, while the interevent time $\tau$ represents the starvation time during which agents do not collect any targets. The burstiness parameter is quantified by $B \equiv (\sigma_\tau-m_\tau)/(\sigma_\tau+m_\tau)$, where $m_\tau$ and $\tau_m$ are the mean and standard deviation of $\tau$, respectively. Fig. \ref{fig:3}\textit{C} shows the PDFs of interevent times, where cases with high and low $\rho$ exhibit a larger proportion of starvation time, reflecting greater burstiness and leading to lower efficiency. The inset in Fig. \ref{fig:3}\textit{C} plots the burstiness parameter $B$ as a function of $\rho$, where $\rho$ values with lower $B$ correspond to the optimal $\rho$. Thus, the analysis of the time series and the derived burstiness provides another perspective on the optimal $\rho$, which not only exhibits the highest efficiency but also steady intake of resources and reduced burstiness in starvation time.

In addition, we measure individual efficiency and examine its variation within a group of agents. Using the same parameter values, we plot the PDFs of the number of targets collected by each agent, as shown in Fig. \ref{fig:3}\textit{D}. From the low-$\rho$ regime to the high-$\rho$ regime, the shape of PDF becomes flatter with increasing probability of medium $N$ values, and the range of the number of targets collected by each agent narrows, indicating that the benefits from targets are distributed more equally among agents. However, while a larger $\rho$ enhances both equity and efficiency, an excessively large $\rho$ increases equity but reduces efficiency and increases burstiness. {These results address RQ2: increasing $\rho$ simultaneously increases equity but can destabilize target intake dynamics, and the $\rho$ that maximizes mean efficiency also minimizes starvation burstiness.}

\begin{figure*}[h!]%[tbhp]
\centering
\includegraphics[width=1\linewidth]{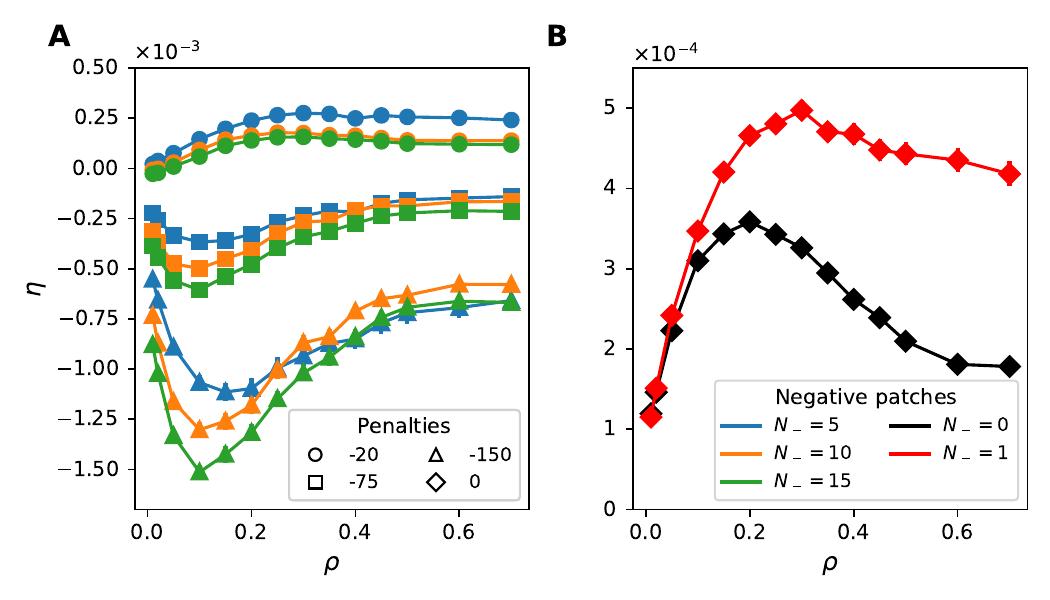}
\caption{\textbf{Negative patches changes optimal $\rho$ and introducing fake negative target can improve efficiency.} (\textit{A}) Efficiency as a function of $\rho$ for three penalty coefficients ($P=-20$, $P=-75$, and $P=-200$). Colors indicate the number of negative patches ($N_-=5$, $10$, and $15$), while the number of positive patches is fixed at $N_+=10$. The domain contains 6000 positive targets and 6000 negative targets; the target distribution is controlled by $\beta=2$, and the group size is 50. As $|P|$ increases, the efficiency-maximizing $\rho$ shifts toward larger values, indicating that broader information sharing becomes more beneficial when negative encounters carry larger penalties. (\textit{B}) When the penalty is ignored ($P=0$), introducing a single negative patch ($N_-=1$) yields higher efficiency than the all-positive baseline ($N_-=0$), showing that randomly placed, uninformative negative cues can improve efficiency by constraining exploration away from unproductive regions.
\label{fig:4}} 
\end{figure*}

\subsubsection*{Social Network Analysis}
Considering that agents utilize social learning to collaborate and improve search efficiency, it is natural to use network metrics to quantify the collective dynamics of our model. To define the social network, we assume that an edge exists between two agents when their mutual distance is smaller than the social learning range $\rho$. We then extract the time-dependent network structure throughout each realization and compute the corresponding network statistics (see Supplementary Materials for details).

Among these metrics, a component is defined as a maximal set of nodes such that every pair of nodes in the set is connected by at least one path. As shown in Fig. \ref{fig:3}\textit{E} and \textit{F}, when $\rho=0.01$, the PDF of the average component size per case $S$, peaks at $S=1$, while the PDF of the average number of components per case $N_C$, peaks at $N_C=50$. This indicates that agents search largely independently in most realizations. As $\rho$ increases, the PDF of $S$ shifts to the right, indicating the formation of larger connected components. Correspondingly, the number of components decreases, leading to a leftward shift of the PDF of $N_C$. When $\rho$ increases to $0.7$, the PDF of $S$ peaks at $S=50$ and the PDF of $N_C$ peaks at $N_C=1$, indicating that agents typically form a single, fully connected cluster. 

The insets in Fig.~\ref{fig:3}\textit{E} and \textit{F} further show that the case-averaged component size, $\langle S \rangle$, increases monotonically with $\rho$, while the case-averaged number of components, $\langle N_C \rangle$, decreases monotonically with $\rho$. Notably, the parameter value corresponding to optimal search efficiency occurs at an intermediate connectivity level, with $\langle N_C \rangle = 7.44$ and $\langle S \rangle = 7.73$.

The component-based network statistics in Fig. \ref{fig:3}\textit{E} and \textit{F} provide a direct network-level interpretation of the social learning range $\rho$. As $\rho$ increases, the implicit proximity-based social network transitions from a highly fragmented structure to a densely connected one. For small $\rho$, the network consists of many isolated components, indicating weak connectivity and limited information transmission between agents. As $\rho$ increases, components merge and the average component size $\langle S\rangle$ increases monotonically, while the average number of components $\langle N_C\rangle$ decreases. In the large-$\rho$ limit, agents typically form a single giant component, implying near-global information sharing in which information discovered by one agent rapidly propagates throughout the group. Notably, the parameter regime corresponding to maximal search efficiency coincides with an intermediate level of connectivity, where neither extreme fragmentation nor full connectivity dominates. This observation parallels results from collective decision-making models on social networks, which show that moderate connectivity often optimizes group performance by balancing information integration and independence \cite{hahn2020truth}, and contrasts with regimes of excessive connectivity that can suppress independent discovery \cite{mann2022collective}.

Beyond static connectivity, the same component statistics also reveal a dynamical mechanism consistent with the emergence of transient diversity. At low $\rho$, the persistence of many small components reflects sustained heterogeneity in agent interactions and information states; however, limited connectivity prevents efficient coordination. At high $\rho$, rapid formation of a single connected component leads to fast homogenization of information and behavior, resulting in premature convergence and redundant exploitation. In contrast, at intermediate $\rho$, the network exhibits a transiently fragmented structure: multiple components coexist during early exploration and progressively merge as information accumulates. This regime maintains heterogeneity long enough to support effective exploration while still allowing coordinated exploitation at later stages. In this sense, $\rho$ regulates not only the degree of network connectivity but also the timescale over which diversity persists. The observed efficiency maximum therefore coincides with a regime in which diversity is generated and maintained transiently rather than permanently, consistent with the general principle that transient diversity enhances collective problem solving by delaying premature consensus \cite{smaldino2024maintaining}. {Taken together, these results close \textbf{RQ2} by showing that $\rho$ acts as a connectivity control parameter for the emergent proximity network: small $\rho$ yields fragmented interaction graphs and localized information flow, whereas larger $\rho$ rapidly produces a highly connected regime that homogenizes behavior. This connectivity transition mechanistically links the observed increase in equity with $\rho$ to changes in temporal stability, since highly connected networks synchronize exploitation and can amplify bursty intake dynamics.} {Having characterized how $\rho$ shapes efficiency and collective organization in the baseline setting, we next use the same framework in a post-hoc application to examine how the presence of risk in the environment affects collective search dynamics, with agents sharing information to avoid negative targets while maximizing positive target intake.}

\subsubsection*{{Collective foraging in risky environments}} Social learning enables agents to aggregate and collect targets, thereby increasing efficiency. In nature, however, prey foraging for food face multifaceted risks, including predation, environmental hazards, and resource depletion, which can compromise their survival and efficiency. Here we ask how such risks alter the functional role of social learning in collective search. To model this, we introduce negative targets that impose a penalty when encountered, while allowing agents to socially transmit hazard information so that nearby agents can actively avoid these locations after receiving social signals (\textbf{see Methods}). We then examine how risk reshapes the dependence of efficiency on the social learning range $\rho$.

% \ZLnote{Lei: same amount of target in a group?}   
We fixed the total number of positive and negative targets at 6000 each. The number of positive patches, $N_+$, is set to 10. We then vary the number of negative patches, $N_-$, {to tune the spatial structure and coverage of hazardous regions}. To measure the efficiency, we measure the number of collected positive targets ($N_{collect+}$) and encountered negative targets ($N_{collect-}$), accounting for penalties from negative targets, using the formula $\eta = (N_{collect+} + P \cdot N_{collect-}) / (N_{positive} \cdot T)$. Here, $P$ is the penalty coefficient, representing {the penalty incurred per negative target encountered}. {As $|P|$ increases, the efficiency-maximizing $\rho$ shifts toward larger values because socially-shared avoidance information becomes more valuable when negative encounters carry larger penalties. This favors larger $\rho$, which spreads avoidance cues more broadly, although excessively large $\rho$ can also increase redundancy in positive-target exploitation.}%\sout{Because larger penalized coefficients change the major concern for maximizing efficiency, shifts from rapidly collecting more positive targets to avoiding encountering negative targets, which necessitate greater $\rho$, although it will diminish the efficiency contributed from positive targets.}

For the same penalty coefficient $P$, we observe that the differences in efficiency curves between cases with varying numbers of negative patches increase as the penalty rises. {This occurs because increasing $N_-$ changes the spatial coverage of hazardous regions, which increases the likelihood that agents incur penalties (or expend travel effort detouring around hazards), an effect that becomes more pronounced as $|P|$ increases.}

% \sout{From Fig. \ref{fig:4}\textit{B}, When the penalty is ignored ($P=0$), randomly distributed negative patches can increase efficiency. \sout{This occurs because detected negative targets from randomly generated patches exclude certain regions of the domain, thereby constraining the area that agents explore, which reduces the time required to explore a new patch.} \ZLnote{This occurs because socially shared negative cues act as exclusion signals that prune unproductive regions of the search space, reducing exploration time before locating a new positive patch.} \ZLnote{Overall, this post-hoc application shows that when negative encounters carry penalties, broader information sharing shifts from merely improving exploitation to actively supporting hazard avoidance, thereby pushing the efficiency-optimal $\rho$ toward larger values. In the limiting case $P=0$, negative cues can still improve performance by acting as exclusion signals that prune the explored domain, illustrating that social information about ``where not to search'' can reshape collective efficiency even when it carries no direct payoff.
% }}

{When the penalty is ignored ($P=0$; Fig.~\ref{fig:4}\textit{B}), randomly distributed negative patches can still increase efficiency. In this limit, socially shared negative cues act as exclusion signals that prune the search space, reducing exploration time before locating a new positive patch. Overall, this post-hoc application shows that when negative encounters carry penalties, broader information sharing shifts from primarily improving exploitation to actively supporting hazard avoidance, thereby pushing the efficiency-optimal $\rho$ toward larger values; when penalties are absent, ``where-not-to-search'' information can still improve performance by restricting exploration away from unproductive regions.}

% This is true even though the patch centers are independently generated. The improvement in efficiency from adding negative patches does not appear to depend on the number of negative patches. 
%\ZLnote{Lei: what does this mean?}While a higher number of negative patches may lead to more negative targets being detected, this has little effect on the overall efficiency of the system. 

\section*{Discussion}
%\sout{The presented work advances the understanding of collective foraging by demonstrating that the social learning range $\rho$—the distance over which individuals respond to social signals—acts as a crucial tuning parameter that regulates a complex interplay between efficiency ($\eta$), temporal stability, and equitable resource distribution in group foraging. Building on the network-level and dynamical analyses presented above, where the social learning range $\rho$ was shown to regulate both network connectivity and the temporal persistence of heterogeneity. Our central finding is that optimal foraging efficiency occurs at intermediate values of $\rho$, offering a mechanistic explanation for patterns observed across disparate biological and computational systems} \cite{martinez2013optimizing,bhattacharya2014collective,garg2022individual,garg2024evolution}. \sout{This work advances beyond previous literature by explicitly quantifying the multi-faceted costs and benefits of social information sharing in dynamic and risky environments. From a broader network perspective, our results place the social learning range $\rho$ within a class of collective systems whose performance depends non-monotonically on effective connectivity. Similar to findings in abstract social networks, intermediate connectivity balances information integration and independent exploration, whereas overly sparse or overly dense networks degrade collective performance} \cite{hahn2020truth, mann2022collective}.

{The present work advances the understanding of collective foraging by showing that the social learning range $\rho$---the distance over which individuals respond to social signals---acts as a key control parameter that jointly regulates foraging efficiency ($\eta$), temporal stability, and equity of resource acquisition. Building on the dynamical and network-level analyses above, we show that $\rho$ tunes both proximity-network connectivity and the persistence of behavioral heterogeneity through time. Our central finding is that efficiency is maximized at intermediate $\rho$, providing a mechanistic explanation for non-monotonic performance patterns reported across diverse biological and computational systems} \cite{martinez2013optimizing,bhattacharya2014collective,garg2022individual,garg2024evolution}. {Beyond identifying this optimum, we explicitly quantify the coupled benefits and costs of social information sharing in environments with depleting positive targets and risky negative targets, clarifying when additional connectivity promotes coordination versus when it induces redundancy and suppresses independent discovery. From a broader network perspective, our results place range-limited social learning within a class of collective systems in which performance depends non-monotonically on effective connectivity: intermediate connectivity balances information integration with continued independent exploration, whereas overly sparse or overly dense networks degrade collective performance} \cite{hahn2020truth,mann2022collective}.

Our finding of an optimal intermediate range for information sharing confirms a recurring theme in the collective search and foraging literature. The peak efficiency observed at intermediate $\rho$ is consistent with the theoretical finding in Mongolian gazelles \cite{martinez2013optimizing}, where acoustic communication over intermediate distance optimizes search efficiency. This pattern of optimal intermediate communication also provides a mechanistic explanation for Deborah Gordon’s work on harvester ants, where restrained signaling maintains efficient colony foraging \cite{gordon1996organization,gordon2013rewards,gordon2014ecology}. A similar mechanism is observed in organizational innovation networks, where excessive connectivity among inventors causes premature convergence on suboptimal solutions; placing less-imitative inventors at influential network positions helps maintain exploratory diversity and improves performance in complex landscapes \cite{heydari2025core}. Our framework generalizes these empirical observations by explicitly linking $\rho$ to the balance of three distinct behavioral modes: global exploration ($\mu=1.1$), local exploitation ($\mu=3$), and targeted walk. Furthermore, the model offers a clearer mechanistic explanation for why non-optimal range fails. When $\rho$ is too small, agents search nearly independently, leading to maximum exploration but minimum cooperation, akin to a purely individual search. This is consistent with finding that insufficient information sharing hampers successful searches \cite{couzin2005effective, torney2009context, rosenthal2015revealing}. On the other hand, if $\rho$ is too large, excessive targeted walks occur, leading to redundant search effort and premature depletion of local patches. This highlights a cost of social learning that suppresses the crucial simultaneous discovery of multiple patches \cite{bhattacharya2014collective}, an effect observed in empirical studies of bird foraging \cite{ward2008quorum, biro2016bringing}.

Our model incorporates adaptive switches between two L\`evy walk modes ($\mu=1.1$ and $\mu=3$), providing a composite search strategy that advances beyond traditional L\`evy models. Pure, memoryless L\`evy walk , while optimal for sparse, randomly distributed targets \cite{viswanathan1999optimizing, sims2008scaling, humphries2010environmental}, often fail to account for the adaptive behavior of real foragers. Our model supports the notion that encounter-conditional heuristics, such as Area-Restricted Search (ARS) triggered by a successful encounter, are often superior to idealized L\`evy walks, especially in patch environments \cite{ross2018general}. Moreover, the targeted walk mode, absent from individual ARS models, serves as a socially-mediated exploitation mechanism. Our analysis of the targeted walk's benefit-to-cost ratio provides a quantitative link between a social behavior (joining successful peers) and the resulting gain in efficiency, clearly demonstrating that the highest benefit-to-cost ratio coincides with the optimal $\rho$.

Prior theoretical studies on collective foraging primarily focused on maximizing search efficiency \cite{torney2009context,torney2011signalling,bhattacharya2014collective,garg2022individual,garg2024evolution}. Our work provides a missing, multifaceted perspective by explicitly quantifying the trade-offs $\rho$ creates with temporal stability and equity of resource intake. The impact of $\rho$ on the burstiness parameter and instantaneous efficiency reveals a crucial trade-off between maximizing average efficiency and maintaining a steady resource stream. While high $\rho$ can lead to short, intense bursts of instantaneous efficiency, it ultimately results in a high burstiness and increased starvation time. This bursty, highly unstable dynamics is consistent with ``herding" and maladaptive informational cascades observed in human social learning experiments \cite{toyokawa2019social}. The lowest burstiness (maximal stability) is found at the same intermediate $\rho$ that optimizes $\eta$. This suggests that optimal efficiency is intrinsically linked to stability in dynamic foraging environments. Social learning is a ``double-edged sword'' , but our model shows the optimal range manages this risk, mitigating the tendency toward rapid, inflexible, and ultimately unsustainable resource exploitation that plagues systems with high social dependence \cite{toyokawa2019social}. Taken together, our results demonstrate a concrete realization of transient diversity—the temporary coexistence of heterogeneous behaviors and information states, which emerges naturally at intermediate $\rho$. Intermediate $\rho$ maintains transient diversity long enough to explore alternative solutions before consensus, echoing a general principle found across formal models of collective problem solving \cite{smaldino2024maintaining}.

By measuring the distribution of collected targets per agent, our work directly addresses the producer-scrounger paradigm from a spatial and kinematic perspective \cite{bhattacharya2014collective,dumke2016producers}. Our results demonstrate that increasing $\rho$ flattens the distribution of collected targets among agents, thereby increasing equity. However, this gain in equity at high $\rho$ comes at the cost of overall lower efficiency and increased instability, as the collective engages in redundant search efforts. Our findings provide an individual-level mechanism for the group-size effects seen in real systems: In the \textit{Australomisidia ergandros} spider, larger group sizes increase scrounger-type frequency and decrease producer frequency \cite{dumke2016producers}. For human social learning, rates of copying increase with group size \cite{toyokawa2019social}. These experimental observations have been 
verified by our model. The time ratio of the targeted walk reaches 0.6 in a group of 50 agents, which is significantly higher than 0.4 in a group of 10 agents, as shown in Supplementary Figure \ref{Supplfig:5}, with the same target distribution.

Our incorporation of negative targets (risks) into the collective search domain introduces a novel dimension of realism to foraging models. The optimal $\rho$ adpatively shift from intermediate to higher as the penalty magnitude increases. This is a key insight: when risks are severe, the group's priority shifts from maximizing immediate resource collection to prioritizing risk avoidance through wider social information gathering. Based on that, we surprisingly found that adding randomly distributed ``fake" negative patches (with zero penalty) significantly increased overall efficiency. This novel emergent mechanism arises because the negative social signals effectively constrain the search space, focusing the collective exploration effort away from empty regions, leading to faster patch discovery. This constraint operates analogously to an informational boundary, pruning uninformative regions while preserving sufficient behavioral diversity to avoid premature convergence. This suggests a completely new principle for designing cooperative search algorithms in robotics and swarms \cite{garnier2007biological,romanczuk2012active,torney2011signalling}.

{The negative-target extension should be viewed as a post-hoc application of the core framework rather than a separate research question. It illustrates how the same control parameter $\rho$ can change its functional role under additional task constraints: when hazards impose penalties, increasing $\rho$ provides value by spreading avoidance information and reducing costly encounters; when penalties are absent, negative cues can still improve efficiency by pruning unproductive regions of the search space. This framing suggests that range-limited social learning can be used diagnostically to anticipate how optimal communication scales shift when search is coupled to risk.}

In summary, this work presents a unified mechanistic framework that integrates efficiency, stability, equity, and risk sensitivity within collective search. By demonstrating that these competing objectives are simultaneously optimized at intermediate social learning ranges, we reveal a general principle underlying collective adaptation. The results reconcile diverse empirical observations—from ants and gazelles to human groups—and provide a testable foundation for designing resilient, self-organized systems in both biological and engineered collectives.

%% file: sections/matmethods.tex
\subsection*{Search space and target distribution} The search space is represented by a two-dimensional grid $(1000\times 1000)$ on a square with a unit length side $(d=1)$, so that each cell of the grid has dimensions $d_{min}\times d_{min}$, where $d_{min} = 10^{-3}$. The total number of targets $N_T$, is grouped into $N_+$ positive target patches or $N_-$ negative patches. The patches are initialized at $N_++N_-$ random locations, and additional targets are placed around previously placed resources at distance $d_T$, according to the probability of $P(d_T) = C d_T^{-\beta}$, where $d_{min}\leq d_T \leq L$, and $C$ is the normalization constant that ensures the probability distribution sums to 1, $C=(1-\beta)/(L^{1-\beta}-(d_{min})^{1-\beta})$. $\beta$ controls the pattern of clustering of targets, where $\beta\rightarrow 1$ resembles a uniform distribution, and $\beta\rightarrow 3$ generates a tightly clustered target distribution. The resources in the model are destructive; that is, they are removed from the environment after being detected by agents. 

\subsection*{Lévy walks with exploration and exploitation modes} The individual search strategy of the agents is based on the Lévy walk, in which the moving orientation of the agents is chosen randomly and the distance (d) is determined by the following distribution, $P(d) = Cd^{-\mu}$, where $d_{min}\leq d \leq L$, and $\mu$ is the power-law exponent. $C$ is a normalization constant, $C=(1-\mu)/(L^{1-\mu}-(d_{min})^{1-\mu})$. The Lévy exponent $\mu$ modulates the individual search behavior between shorter steps and longer steps. As in $\mu\rightarrow 1$, the agents move in a longer step with an explorative strategy, while $\mu\rightarrow 3$. During each timestep, agents move in fixed step sizes of $d_{min} = 10^{-3}$ to cover a total step length $d$, which is sampled from a probability distribution. Each step corresponds to a constant speed of $d_{min}$ per timestep, and larger values of $d$ require more time steps to complete. At each timestep, each agent searches within a radius $r = d_{min}$ and collects one target if it exists in the range.

\subsection*{Effects of detected negative target}
When an agent detects a negative target, it shares the target's location with other agents within a radius of $\rho$ around itself. The interaction between a detected target and agents aware of its location is modeled using an exponentially decaying function, $l = 0.002e^{-50d}$, where $d$ represents the distance between the agent and the detected target, and $l$ is the repulsive step length directed outward from the detected target to the agent's current position. This function decays rapidly, and $l$ falls below the minimum step length $d_{min}$ when $d > R$, ensuring that $l$ has minimal effect on agents far from the detected target and allows agents to avoid the region around the negative target within a radius $R$. This repulsive step is superimposed on the agents' movement, which follows a Lévy walk or targeted walk, thereby altering the agents' moving direction while maintaining their constant speed of $d_{min}$ per timestep.

\subsection*{Simulation setup and numerical implementations} 
{Because targets are finite and non-regenerative, efficiency necessarily decreases as depletion progresses, making raw time-to-termination comparisons across conditions sensitive to the depletion level. To enable consistent comparisons across different $\rho$, we therefore evaluate performance at a standardized depletion condition and terminate each realization once agents have collected $30\%$ of the positive targets \cite{garg2022individual,garg2024evolution}. To verify that our qualitative conclusions are not an artifact of this stopping rule, we also examine later-stage depletion by running simulations beyond the $30\%$ threshold in settings with fewer patches; these results are shown in Supplementary Fig. \ref{Supplfig:4}.} Considering that our model is based on a random walk, which introduces randomness to the statistics, we use the same set of target distributions for cases with different $\rho$ to reduce variance. To minimize the errorbar with a 95\% confidence interval, we run 2000 cases for each $\rho$.

\section*{Code and Data availability}
All python scripts used to generate the figures and reproduce the simulation results in this paper are openly available on our GitHub repository: \\ https://github.com/LoneStar97/social-learning-search.git  \\ Additional videos of agents' movement can be find here: \\ https://www.youtube.com/playlist?list=PLgRFM9nAjJRwoZvCGBAdCIE-BYNgPmSuV

%% file: sections/acknowledgement.tex
Research was sponsored by the DEVCOM Analysis Center and was accomplished under Contract Number W911QX-23-D0002. The views and conclusions contained in this document are those of the authors and should not be interpreted as representing the official policies, either expressed or implied, of the DEVCOM Analysis Center or the U.S. Government. The U.S. Government is authorized to reproduce and distribute reprints for Government purposes, notwithstanding any copyright notation herein.

This research was supported in part by the University of Pittsburgh Center for Research Computing, RRID:SCR\_022735, through the resources provided. Specifically, this work used the HTC cluster, which is supported by NIH award number S10OD028483.

We thank Janson W. Burton, Lingfei Wu, and Hirokazu Shirado for helpful discussion.

%% file: sections/si.tex
\newpage
\setcounter{page}{1}
\setcounter{equation}{0}
\setcounter{figure}{0}
\renewcommand{\thepage}{S\arabic{page}}
\renewcommand{\thesection}{S\arabic{section}}
\renewcommand{\theequation}{S.\arabic{equation}}
\renewcommand{\figurename}{Supplementary Figure}
\renewcommand{\tablename}{Supplementary Table}
% To modify the text used at the start of a caption:
% \renewcommand{\figurename}{Supplemental Material, Figure}
\noindent {\Huge \bf Supplementary Information} \\
\\
This Supplementary Information provides additional analyses and robustness checks that support the main text. Section S1 examines how the target-clustering parameter $\beta$ affects search efficiency and behavioral time allocations. Section S2 reports time-resolved (instantaneous) search efficiency to clarify how target depletion influences the efficiency curves and the optimal social learning range $\rho$. Section S3 evaluates the effect of group size on the existence and location of the optimal 
$\rho$. Finally, Section S4 details the construction of the proximity-based social network induced by social learning and defines the network metrics used to quantify connectivity and the emergence of collective structure over time.

\noindent {\small
\tableofcontents}
% \section{Additional Related Work}

\section{Target distribution}
The parameter $\beta$ controls the spatial distribution of targets around the target seeds, such that $\beta \to 1$ results in a nearly uniform distribution, while $\beta \to 3$ produces a distribution with tightly clustered targets. The main text presents the results using a target distribution with $\beta = 2$. Here, we show results for 50 agents with $\beta = 1.1$ and $\beta = 3$, with the number of patches set to 10, 20, and 50, consistent with the main text.

\begin{figure*}[h!]%[tbhp]
\centering
\includegraphics[width=1\linewidth]{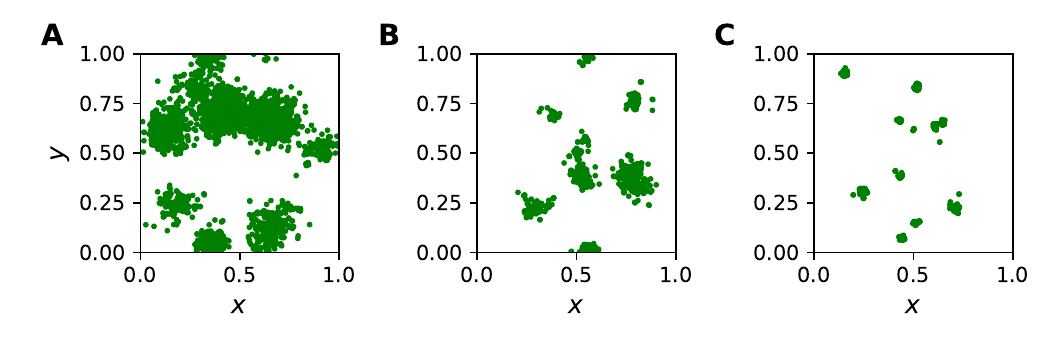}
\caption{\textbf{$\beta$ controls the clustering of targets within a patch} (\textit{A})(\textit{B})(\textit{C}) The target distributions feature 10 patches of 600 targets each, with varying degrees of clustering controlled by $\beta=1.1$, $\beta=2$, and $\beta=3$.
\label{Supplfig:1}} 
\end{figure*}

\begin{figure*}[h!]%[tbhp]
\centering
\includegraphics[width=1\linewidth]{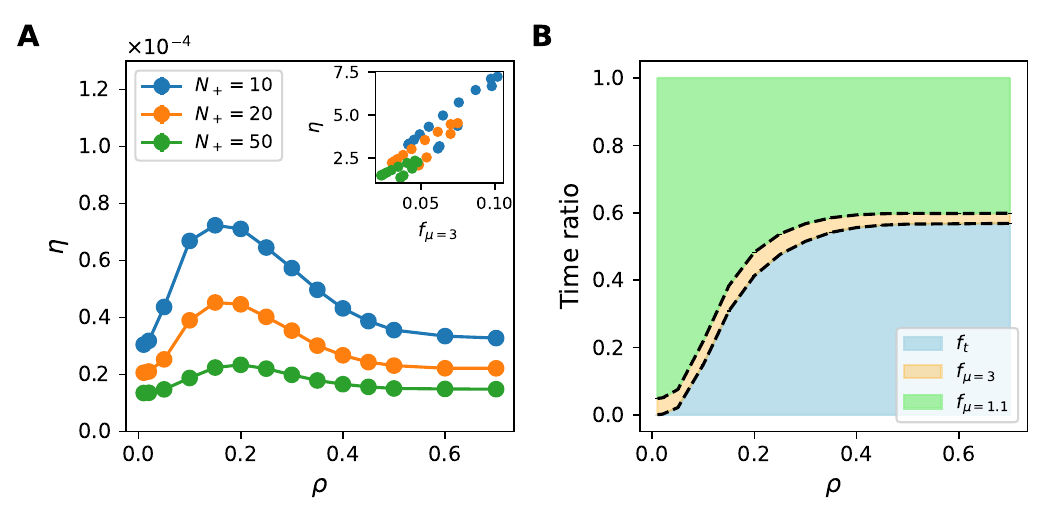}
\caption{\textbf{Sparsely distributed targets per patch lowers the efficiency and time ratio of exploitation} (\textit{A}) Plots of efficiency as a function of $\rho$ with a fixed number of agents $N_A$ = 50 and a series of number of patches, $N_+$ = 10, $N_+$ = 20, and $N_+$ = 50, while the total number of targets within the domain is fixed at 6000. The controlling parameter of the target distribution is $\beta=1.1$. Error bars indicating 95\% confidence interval are too short to show on the figure. The inset shows that efficiency is positively correlated with $f_{\mu=3}$, time ratio of exploitation. (\textit{B}) Averaged time ratios spending on three behaviors for the case of $N_+=20$ in (\textit{A}) with 50 agents.
\label{Supplfig:2}} 
\end{figure*}

\begin{figure*}[h!]%[tbhp]
\centering
\includegraphics[width=1\linewidth]{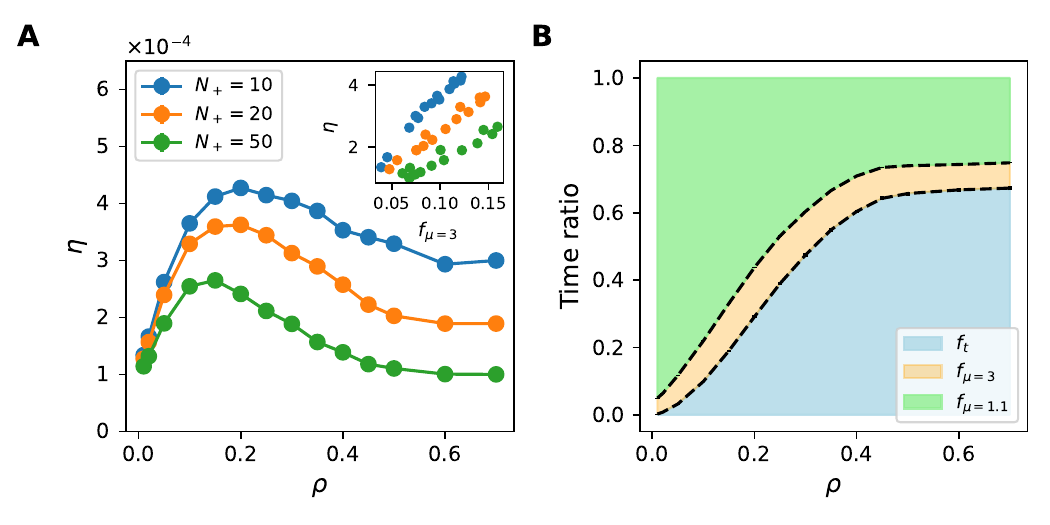}
\caption{\textbf{Closely distributed targets per patch increases the efficiency and lower the time ratio of exploitation} (\textit{A}) Plots of efficiency as a function of $\rho$ with a fixed number of agents $N_A$ = 50 and a series of number of patches, $N_+$ = 10, $N_+$ = 20, and $N_+$ = 50, while the total number of targets within the domain is fixed at 6000. The controlling parameter of the target distribution is $\beta=3$. Error bars indicating 95\% confidence interval are too short to show on the figure. The inset shows that efficiency is positively correlated with $f_{\mu=3}$, time ratio of exploitation. (\textit{B}) Averaged time ratios spending on three behaviors for the case of $N_+=20$ in (\textit{A}) with 50 agents.
\label{Supplfig:3}} 
\end{figure*}

In the main text, the optimal $\rho$ decreases as $N_+$ increases when $\beta = 2$, because exploring a new patch becomes more important than exploiting a known one in this scenario. This effect is less pronounced when $\beta = 1.1$, as targets are scattered more extensively around each patch center. In this scenario, targets are so sparsely distributed that agents need to rely on exploration even within the same patch to collect targets. As a result, the time ratio of exploitation is minimal, while that of exploration increases compared to the case where $\beta=2$. Since agents already are highly dependent on exploration, the optimal $\rho$ does not decrease significantly as $N_+$ increases, as shown in Supplementary Fig. \ref{Supplfig:2}. Conversely, this effect strengthens when $\beta = 3$, where targets are closely clustered around each patch center, making agents more dependent on exploration to find targets in a new patch. Therefore, the optimal $\rho$ decreases significantly as $N_+$ increases, as shown in Supplementary Fig. \ref{Supplfig:3}. For cases with the same number of patches $N_+$, the peak efficiency at the optimal $\rho$ increases as $\beta$ rises from 1.1 to 3. This is because a higher $\beta$ leads to more tightly clustered targets, which reduces the time needed for agents to deplete a patch of its resources. The efficiency curves with different $\beta$ values show that our model is more effective for heterogeneously distributed targets than for randomly distributed targets.

\section{Instantaneous search efficiency over the time}
Considering the targets can be consumed by agents, the efficiency varies during the search with the number of targets decreased over the time. To show the trends of instantaneous efficiency during the search, we compare the efficiency of cases with the same number of targets in each patch as shown in Supplementary Fig. \ref{Supplfig:4}. From this figure, we can learn that the normalized efficiency increases due to the decrease in the number of patches. The optimal $\rho$ increases as the number of patches decreases, meaning that agents need more cooperation through social learning as resources become sparse. The optimal $\rho$ still has the maximal time ratios of exploitation in each case. Moreover, the time ratios of exploitation decrease, meaning that exploration becomes more important as the number of patches decreases.

\begin{figure*}[h!]%[tbhp]
\centering
\includegraphics[width=1\linewidth]{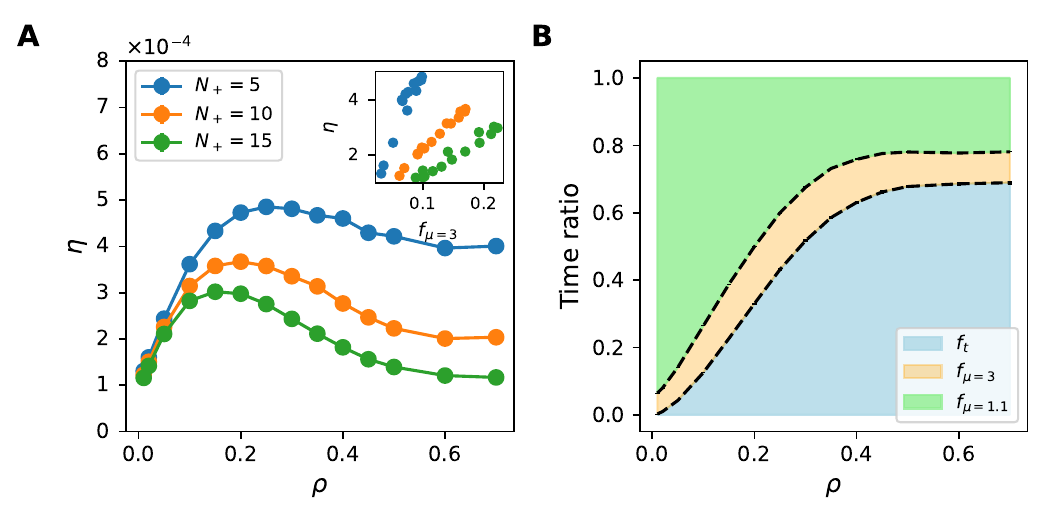}
\caption{\textbf{The optimal $\rho$ increases as the number of patches decreases} (\textit{A}) Plots of efficiency as a function of $\rho$ with a fixed number of agents $N_A$ = 50 and a series of number of patches, $N_+$ = 5, $N_+$ = 10, and $N_+$ = 15, while the number of targets in each patch is fixed and the total numbers of targets are 2500, 5000, and 7500, respectively. The controlling parameter of the target distribution is $\beta=2$. Error bars indicating 95\% confidence interval are too short to show on the figure. The inset shows that efficiency is positively correlated with $f_{\mu=3}$, time ratio of exploitation. (\textit{B}) Averaged time ratios spending on three behaviors for the case of  $N_+=10$ in (\textit{A}) with 50 agents.
\label{Supplfig:4}} 
\end{figure*}

\section{Group size}
After testing the case of 10 agents searching for the targets with the same distribution as that in Fig. \ref{fig:2}, we can see that the optimal $\rho$ still exists, and it corresponds to the maximal time ratios of exploitation. However, the shape of efficiency curves are flatter compared to cases with 50 agents in the main text. This occurs because the smaller number of agents limits the maximum instantaneous efficiency that can be achieved by social learning and the following targeted walk, thereby diminishing the advantages of collective search. As a result, the efficiency curves the efficiency curves show less variation and are less dependent on $\rho$. This suggests that the model is more effective for larger groups. The group size constraint on instantaneous efficiency also leads agents to use a larger $\rho$ to attract more peers to achieve higher instantaneous efficiency, which, in turn, increases the optimal $\rho$. 

\begin{figure*}[h!]%[tbhp]
\centering
\includegraphics[width=1\linewidth]{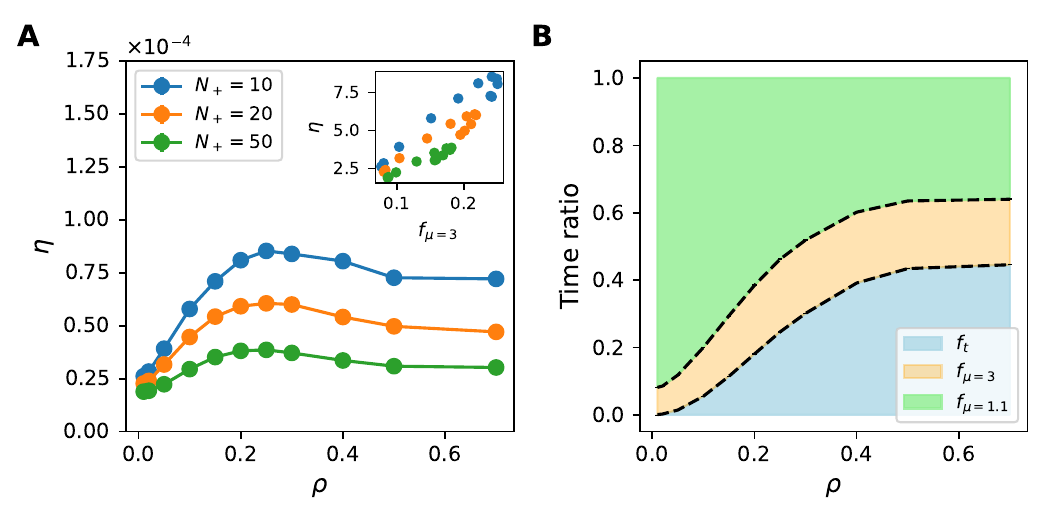}
\caption{\textbf{Smaller group size limited the function of social learning which make efficiency curves fatter} (\textit{A}) Plots of efficiency as a function of $\rho$ with a fixed number of agents $N_A$ = 10 and a series of number of patches, $N_+$ = 10, $N_+$ = 20, and $N_+$ = 50, while the total number of targets within the domain is fixed at 6000. The controlling parameter of the target distribution is $\beta=2$. Error bars indicating 95\% confidence interval are too short to show on the figure. The inset shows that efficiency is positively correlated with $f_{\mu=3}$, time ratio of exploitation. (\textit{B}) Averaged time ratios spending on three behaviors for the case of  $N_+=20$ in (\textit{A}).
\label{Supplfig:5}} 
\end{figure*}

\section{Network Metrics}
To quantify how the social learning range $\rho$ shapes interaction structure, we define an implicit, time-dependent proximity network among agents. At each timestep $t$, each agent is treated as a node, and an undirected edge is placed between agents $i$ and $j$ if their Euclidean separation satisfies
\begin{equation}
\|\mathbf{x}_i(t)-\mathbf{x}_j(t)\| < \rho .
\end{equation}
This produces a sequence of graphs $G(t)$ over the course of each realization. We compute standard network statistics from $G(t)$ to characterize how information-sharing range maps onto effective connectivity, local transitivity, and fragmentation.

The degree of agent $i$ at time $t$ is $k_i(t)$, the number of neighbors within distance $\rho$. Unless otherwise stated, the degree time series shown in Supplementary Fig. \ref{Supplfig:6}\textit{A} is the median (over realizations) of the network-averaged degree at each timestep.

Local transitivity is quantified using the (local) clustering coefficient $C_i(t)$, defined for nodes with $k_i(t)\ge 2$ as
\begin{equation}
C_i(t) = \frac{2T_i(t)}{k_i(t)\left(k_i(t)-1\right)},
\end{equation}
\begin{figure*}[h!]%[tbhp]
\centering
\includegraphics[width=1\linewidth]{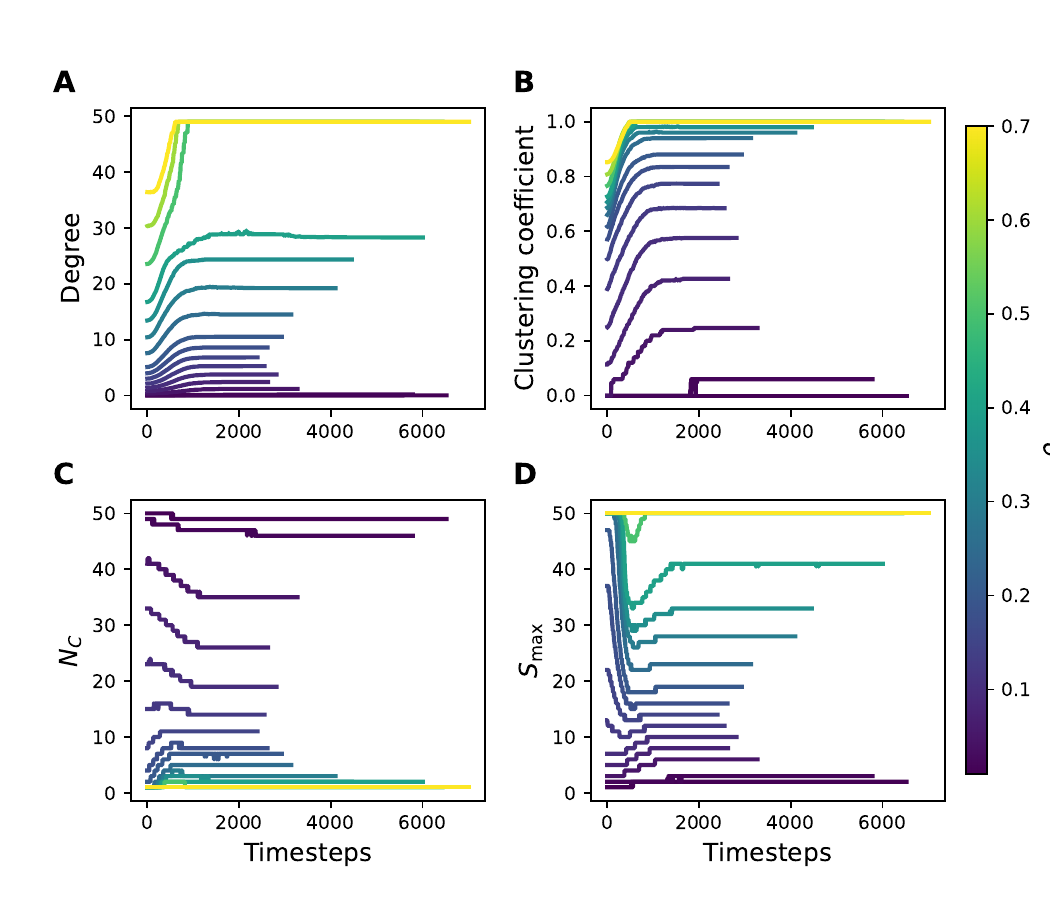}
\caption{\textbf{Network dynamics across social learning ranges.}  Median (over independent realizations) time series of proximity-network metrics for a range of social learning radii $\rho$. At each timestep, an edge is defined between two agents if their separation is smaller than $\rho$. (\textit{A}) Median node degree, showing the growth of effective local connectivity as interactions accumulate and as $\rho$ increases.  Median clustering coefficient, quantifying the emergence of locally transitive interaction neighborhoods. (\textit{C}) Median number of connected components $N_C$, measuring network fragmentation and the transition from many disconnected groups to a single connected network as $\rho$ increases. (\textit{D}) Median largest component size $S_{\mathrm{max}}$, capturing the growth of the giant component and the onset of near-global connectivity at large $\rho$. Together, the panels illustrate how increasing $\rho$ accelerates the collapse of fragmentation and the formation of a highly connected, locally clustered interaction network over time.
\label{Supplfig:6}} 
\end{figure*}

where $T_i(t)$ is the number of triangles that include node $i$. The network clustering coefficient $C(t)$ is taken as the mean of $C_i(t)$ over all nodes with $k_i(t)\ge 2$ (nodes with $k_i<2$ contribute zero by convention). The time series in Supplementary Fig. \ref{Supplfig:6}\textit{B} shows the median (over realizations) of $C(t)$. Values near $0$ indicate weakly transitive neighborhoods, whereas values near $1$ indicate highly transitive local structure.

A connected component is defined as a maximal set of nodes in which every pair is connected by at least one path. Let $N_C(t)$ be the number of connected components at time $t$. We also define the component sizes $\{s_1(t), s_2(t), \dots, s_{N_C(t)}(t)\}$, where each $s_m(t)$ is the number of nodes in component $m$. The largest component size is
\begin{equation}
S_{\max}(t)=\max_m s_m(t).
\end{equation}
Supplementary Fig. \ref{Supplfig:6}\textit{C} reports the median (over realizations) of $N_C(t)$, and Supplementary Fig. \ref{Supplfig:6}\textit{D} reports the median (over realizations) of $S_{\max}(t)$. Together, these quantities capture the transition from a fragmented network (large $N_C$, small $S_{\max}$) at small $\rho$ to near-global connectivity (small $N_C$, large $S_{\max}\approx N_A$) at large $\rho$.

In addition to time-resolved metrics, we compute case-level component summaries used in Fig.\ref{fig:3}~\textit{E}--\textit{F} of the main text. Specifically, for each realization we compute (i) the average component size $S$ (obtained by averaging component sizes over time), and (ii) the average number of components $N_C$. We then aggregate across realizations to form the PDFs shown in Fig.\ref{fig:3}~\textit{E}--\textit{F} and to compute the case-averaged trends $\langle S\rangle$ and $\langle N_C\rangle$ versus $\rho$. These component-based statistics provide a compact network-level interpretation of $\rho$: increasing $\rho$ merges components, increases typical connected set sizes, and accelerates the emergence of a giant component, thereby increasing effective information transmission.

For each $\rho$, we run 2000 independent realizations. For time-resolved plots (Supplementary Fig. \ref{Supplfig:6}), we compute the relevant network statistic at each timestep in each realization and then report the median across realizations at each time to reduce sensitivity to outliers. This yields robust trajectories of how connectivity and fragmentation develop during the search process.

% \section{Robustness Checks}

% \subsection{Group Size}

% \subsection{Maintaining the same number of targets per patch}

% \subsection{Clustering and Radius of Target Patches}

% \section{Placement of Fake Negative Targets}
%Densities and hypothesis testing